\documentclass[cameraready]{Interspeech}

\title{Are Audio-Language Models Listening? Audio-Specialist Heads for Adaptive Audio Steering}

\author[affiliation={1}]{Neta}{Glazer}
\author[affiliation={2}]{Lenny}{Aharon}
\author[affiliation={1}]{Ethan}{Fetaya}

\address{
    $^1$ Bar-Ilan University, Ramat Gan, Israel\\
    $^2$ Columbia University, New York, NY, USA
}

\email{neta.glazer@biu.ac.il, Lenny.Aharon@columbia.edu}


\usepackage{comment}
\usepackage{amsmath}
\usepackage{amsfonts}
\usepackage{amssymb}
\usepackage{algorithm}
\usepackage{algpseudocode}
\usepackage{algorithm}
\usepackage{algpseudocode} 
\usepackage{amsmath,amsfonts,amssymb}
\usepackage{xcolor}
\usepackage{multirow}
\usepackage[colorinlistoftodos]{todonotes}

\begin{document}

\maketitle


\begin{abstract}
Multimodal large language models can exhibit text dominance, over-relying on linguistic priors instead of grounding predictions in non-text inputs. One example is large audio-language models (LALMs) where decisive audio evidence can be under-utilized even when it contains important information. To address this issue we use mechanistic interpretability to identify a small set of audio-specialist attention heads whose audio attention yields a “listening” signal. We show that this signal increases when audio evidence affects the model’s output, providing an indicator of audio engagement under standard prompting. Leveraging this localization, we construct an audio–silence steering direction and apply an inference-time activation intervention to the final representation, amplifying the model’s audio effect. To demonstrate the utility of this intervention, we show on MMAU that this improves accuracy by up to +8.0 percentage points on two Qwen-based LALMs, without any parameter updates.

\end{abstract}

\section{Introduction}

Large audio-language models (LALMs) couple a pretrained audio encoder with a decoder-based large language model (LLM), enabling instruction-following understanding and reasoning over speech, environmental sounds, and music from natural-language prompts \cite{chu2024qwen2audio,tang2024salmonn,ghosh2025audioflamingo2,vaswani2017attention}. 
Architecturally, common fusion strategies include
(i) projecting audio-encoder representations into the LLM embedding space and inserting them as a short sequence of audio-conditioned pseudo-tokens processed jointly with text via self-attention \cite{chu2023qwen,tang2024salmonn},
and (ii) conditioning the LLM on audio features through added (often gated) cross-attention adaptor layers, rather than interleaving audio tokens with the text stream \cite{ghosh2025audioflamingo2}.

However, processing these mixed token streams within a single backbone trained predominantly on text introduces a critical phenomenon: \emph{text dominance}. Even when non-text modalities are highly informative, models frequently rely disproportionately on linguistic cues. Recent systematic evidence shows that this imbalance is pervasive across modalities, including audio, and stems from factors like fusion design choices and attention dilution caused by non-text token redundancy \cite{wu2025languageoverrules, aharon2025uncertainty}. Related work frames this phenomenon as \emph{language-prior bias}, demonstrating that multimodal outputs are often driven more by the underlying LLM's priors than by the non-text inputs themselves \cite{zhang2024penalization}. Specifically within the audio-language domain, controlled mismatch studies demonstrate that capable LALMs can be dominated by textual prompts, effectively disregarding contradictory audio evidence \cite{wang2025audio_text_disagree,billa2026audio_llms_dont_listen}.

Mechanistic interpretability has become a central framework for analyzing the internal computations of text-only LLMs, aiming to identify localized mechanisms in weights and activations that causally drive model behavior rather than relying on post-hoc rationales \cite{nanda2023progress,elhage2021framework,meng2022locating}.
More recently, this toolkit has begun to extend to multimodal architectures, including LALMs, enabling component-level analyses of how non-text modalities are integrated and where modality-specific failure modes arise \cite{glazer2025beyond,futami2024finding,golovanevsky-etal-2025-vlms,basile2025headpursuit}.
A key part of this paradigm is the use of causal interventions on internal activations (e.g., ablation or activation patching) to test mechanistic hypotheses.
Building on this idea, steering refers to inference-time interventions that modify internal activations to influence how information is processed \cite{turner2024activation}.
A recurring finding is that transformer components, particularly individual attention heads, exhibit stable and specialized computational roles, enabling causal intervention at the head level and motivating activation-based steering approaches that modulate model behavior during inference \cite{elhage2021framework,olsson2022context,voita-etal-2019-analyzing}.




Motivated by the text-dominance failure mode in LALMs and recent progress in mechanistic interpretability for multimodal transformers, we use mechanistic tools to study audio under-utilization in depth.
In particular, we ask whether head-level signals can indicate when the model is engaging with audio, and whether these signals can be used as a practical handle for inference-time steering.

In this work, we make two main contributions. First, we identify a small set of \emph{audio-specialist} attention heads whose attention to audio is predictive of correctness, yielding an instance-level ``listening'' signal. Second, we demonstrate that mechanistic analysis at the component level can provide an actionable handle in audio-language models: using the localization to guide a controlled inference-time activation intervention, we amplify the model’s \emph{audio effect} and improve MMAU performance for two Qwen-based LALMs (Qwen2-Audio-7B \cite{chu2024qwen2audio} and R1-AQA \cite{li2025r1aqa}), without any parameter updates.



\begin{figure*}[t]
  \centering
  \includegraphics[width=\textwidth]{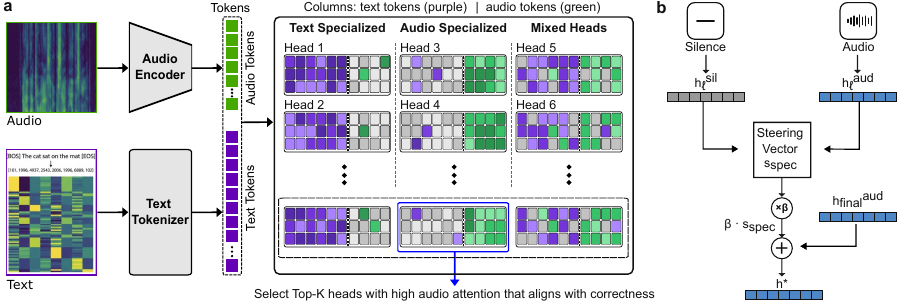}
\caption{\textbf{Specialist-Guided Steering (SGS).}
\textbf{(a)} We identify audio-specialist attention heads by computing each head’s audio-attention share and selecting the Top-$K$ heads whose audio attention is most predictive of correctness on a calibration set.
\textbf{(b)} We run audio and matched-duration silence forward passes and form a layer-localized steering direction by aggregating residual differences $(\mathbf{h}^{\text{aud}}_\ell-\mathbf{h}^{\text{sil}}_\ell)$ over the specialist layer set $\mathcal{L}$ (layers containing the discovered heads). We scale this direction by $\beta$ and add it to the final representation to obtain $\mathbf{h}^*$ for prediction.}
  \label{fig:Specialist_heads}
\end{figure*}

\section{Related Work}
\textbf{Text Dominance in Multi Modal LLMs.} Large audio-language models (LALMs) extend instruction-following LLMs with audio front-ends and multimodal fusion mechanisms, such as token injection, enabling joint reasoning over audio and text \cite{chu2024qwen2audio,ghosh2025audioflamingo2}. However, a reliability concern in these systems is \emph{text dominance} (or language-prior bias), where linguistic cues override informative non-text evidence \cite{wu2025languageoverrules,zheng2025mllms}. This phenomenon has been systematically documented across multimodal settings, where models default to strong language priors even when they conflict with perceptual evidence, leading to spurious correlations, modality under-utilization, and failures to properly ground predictions in the non-text signal \cite{zhang2024penalization,leng2024mitigating, favero2024multi,kuan2025can}.  Within the audio domain, controlled audio-text disagreement studies provide direct evidence of modality arbitration failures, where models prefer textual instructions even when they directly contradict acoustic ground truths \cite{wang2025audio_text_disagree,billa2026audio_llms_dont_listen}. Furthermore, speech affect evaluations show that many LALMs behave more like rigid transcribers than active listeners, failing to disentangle acoustic prosody from lexical content \cite{chen2025audio}.

\textbf{Mechanistic Interpretability.}
Mechanistic interpretability provides tools to localize \emph{where} and \emph{how} information is represented and used within transformer computations, linking model behavior to internal mechanisms rather than post-hoc rationales \cite{wang2022interpretability}.
A recurring finding is that individual components—especially attention heads—often exhibit specialized and reusable functional roles \cite{voita-etal-2019-analyzing}, enabling targeted, component-level interventions.
More recently, these tools have been extended to multimodal transformers, including analyses that identify modality-linked attention heads and study their causal role across tasks \cite{golovanevsky-etal-2025-vlms,basile2025headpursuit,li2023inference}.
Within audio and LALM settings, early work applies mechanistic analyses to probe how acoustic evidence propagates through the model and to localize audio-related computations \cite{glazer2025beyond}.
A complementary line explores training-free, inference-time interventions that exploit such localization to improve multimodal grounding, e.g., vector-steering methods for audio models \cite{turner2024steering,lin2025avs}. A complementary line explores training-free, inference-time interventions that exploit such localization to improve multimodal grounding, including activation steering via adding a learned or contrastive direction to internal representations \cite{rimsky2024steering,turner2023steering}.

\section{Preliminaries and Notation}
\label{sec:prelim}

\noindent\textbf{Token stream and audio indices.}
Audio-language transformers operate on a single sequence of $n$ tokens, including both text and audio tokens.
We denote the set of audio token indices by $\mathcal{I}_{\text{audio}}\subset\{1,\dots,n\}$.

\noindent\textbf{Multi-head self-attention.}
Given input $x$, self-attention produces in layer $\ell$ and head $h$ an attention matrix $\mathbf{A}_{\ell,h}(x)\in\mathbb{R}^{n\times n}$ whose rows sum to one, where $\mathbf{A}_{\ell,h}[i,j](x)$ is the attention weight from query position $i$ to key position $j$.

\noindent\textbf{Audio attention from the final prompt token.}
Let $i_{\text{final}}$ denote the index of the final token in the prompt (the last position before generation). For head $(\ell,h)$, we compute
\begin{equation}
a_{\ell,h}(x) = \sum_{j\in\mathcal{I}_{\text{audio}}}\mathbf{A}_{\ell,h}[i_{\text{final}},j](x).
\label{eq:audio_attention}
\end{equation}
Since attention rows sum to one, $a_{\ell,h}(x)\in[0,1]$ is the fraction of attention from query position $i_{\text{final}}$ directed to audio tokens.

\noindent\textbf{Residual stream states.}
Let $\mathbf{h}_\ell(x)\in\mathbb{R}^{d_{\text{model}}}$ denote the residual-stream representation at position $i_{\text{final}}$ after layer $\ell$, and let $\mathbf{h}_{\text{final}}(x)$ be the final-layer representation at this position.

\noindent\textbf{Audio ablation.}
To isolate the effect of audio, we use a matched-duration silence baseline.
For each example $x$, we define $x^{\text{aud}}$ (original audio) and $x^{\text{sil}}$ (audio replaced by zeros of the same duration), with corresponding residual-stream states $\mathbf{h}^{\text{aud}}_\ell(x)$ and $\mathbf{h}^{\text{sil}}_\ell(x)$.

\noindent\textbf{Steering intervention.}
Given a direction and strength $\beta$, we apply steering by modifying internal activations during the forward pass and then computing predictions with the model's language modeling head.

\begin{figure}[t]
\centering
\includegraphics[width=\linewidth,height=0.48\linewidth,keepaspectratio]{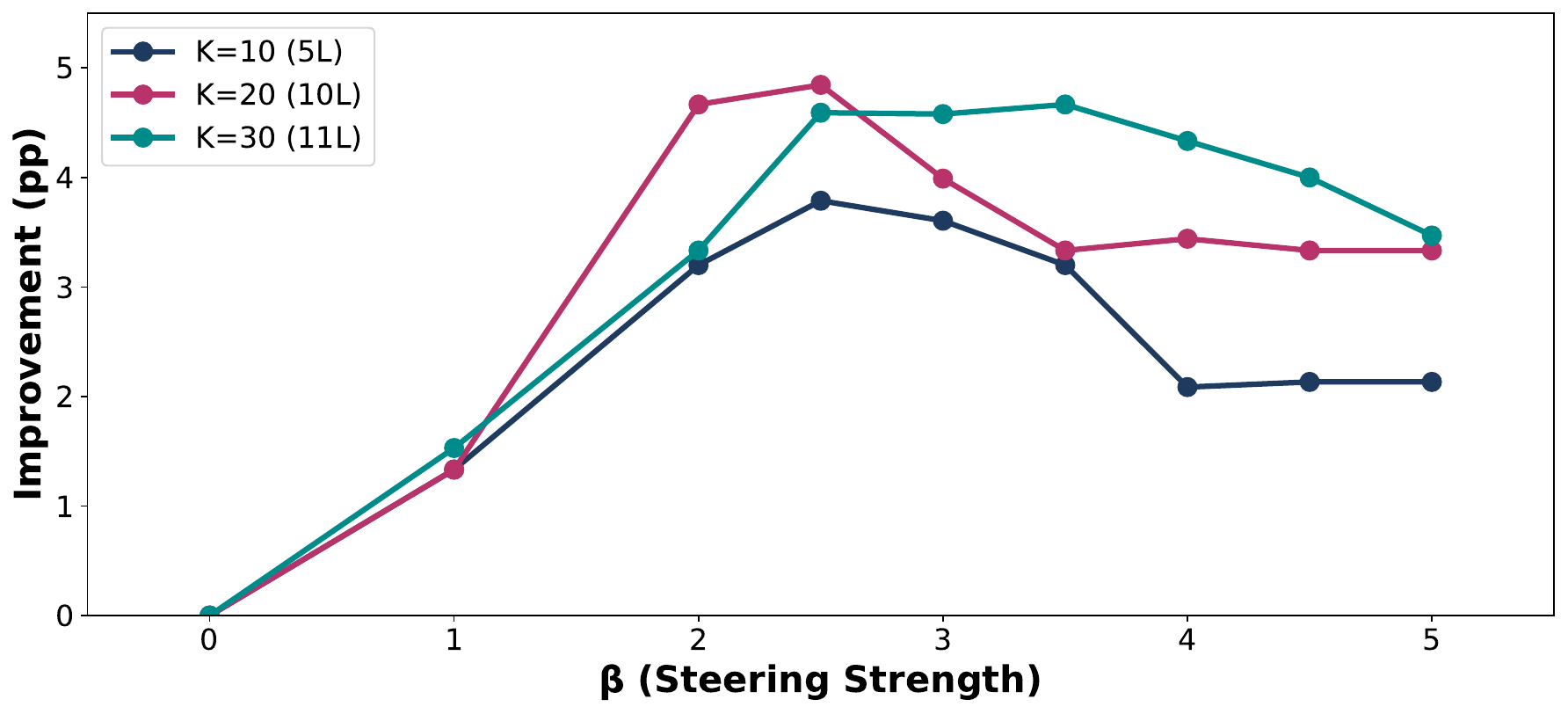}

\vspace{0.6em}

\includegraphics[width=\linewidth,height=0.48\linewidth,keepaspectratio]{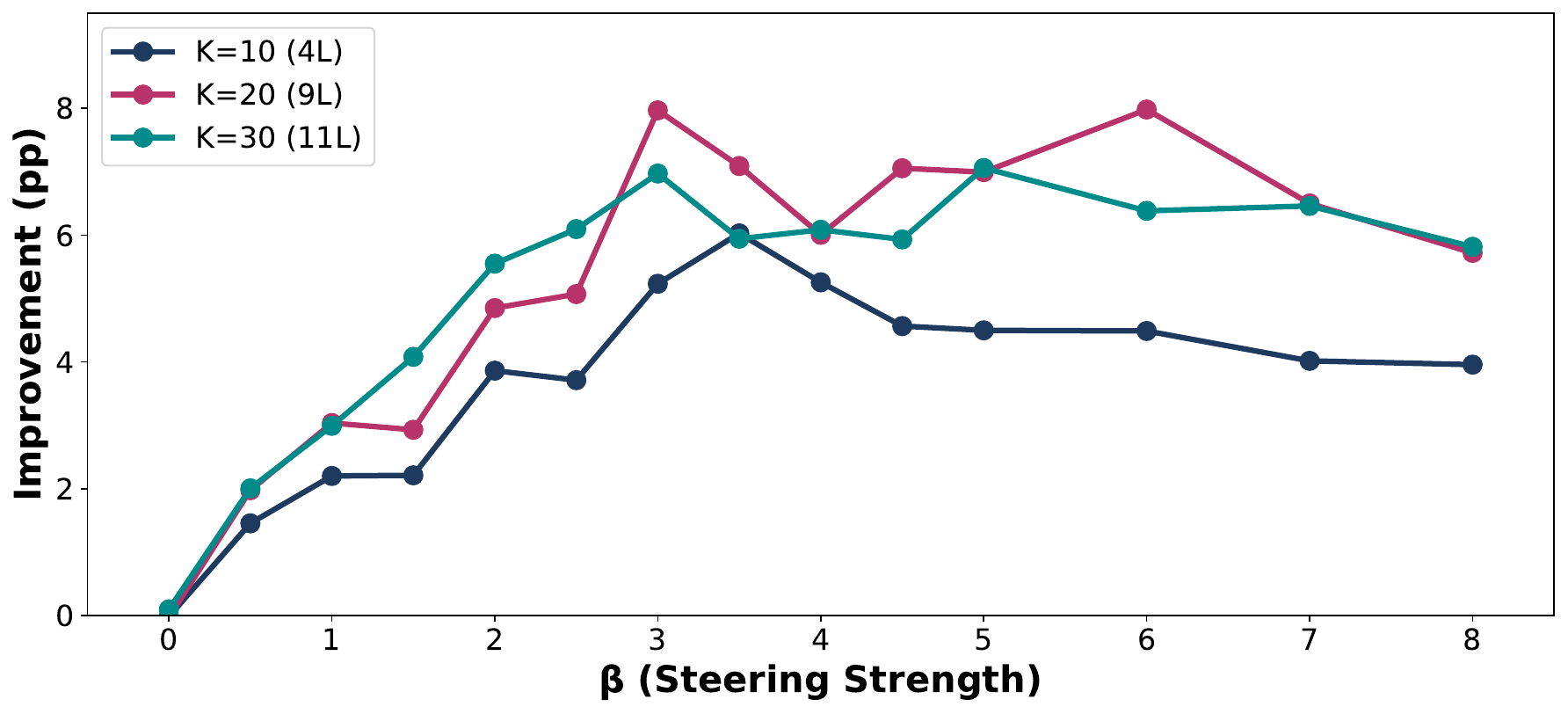}

\caption{Effect of steering strength $\beta$ and specialist count $K$ on performance for R1-AQA (top) and Qwen2-Audio-7B (bottom). Lines show improvement in percentage points (pp) for different Top-$K$ specialist head sets; each $K$ induces a specialist layer set $\mathcal{L}$ and we apply layer-localized steering within $\mathcal{L}$.}
\label{fig:steering_grid}
\end{figure}



\section{Method}
\label{sec:method}

Our approach has two stages. First, we identify a small set of audio-specialist attention heads whose attention to audio is most predictive of correctness on a calibration split, yielding a head-level localization of audio engagement. Second, we use this localization to construct a specialist-constrained audio--silence steering direction and apply a controlled inference-time activation intervention.

\subsection{Discovering Audio-Specialist Heads}
\label{sec:discovery}

\noindent\textbf{Audio attention signal.}
We use the audio attention mass from the final prompt position $i_{\text{final}}$, $a_{\ell,h}(x)$ (Eq.~\ref{eq:audio_attention}), as a head-level measure of audio engagement.

\noindent\textbf{Specialist scoring and selection.}
Using a held-out calibration set $\mathcal{D}_{\text{cal}}$ of multiple-choice questions, we define a binary correctness label
$y(x)=\mathbb{1}[\hat{c}(x)=c^*(x)]\in\{0,1\}$, where $\hat{c}(x)$ is the model's predicted option and $c^*(x)$ is the ground-truth option.
We score each head $(\ell,h)$ by the association between $a_{\ell,h}(x)$ and correctness:
\begin{equation}
\rho_{\ell,h}=\mathrm{corr}\Big(\{a_{\ell,h}(x)\}_{x\in\mathcal{D}_{\text{cal}}},\{y(x)\}_{x\in\mathcal{D}_{\text{cal}}}\Big),
\label{eq:head_corr}
\end{equation}
where $\mathrm{corr}$ denotes Pearson correlation (equivalently, point-biserial correlation for binary $y$).
We define the specialist set $\mathcal{H}_{\text{spec}}$ as the top-$K$ heads ranked by $|\rho_{\ell,h}|$ (we use $K{=}20$).
When forming an instance-level listening score, we aggregate heads with signed ($|\rho|$-weighted) contributions so that heads negatively associated with correctness contribute oppositely.

\noindent\textbf{Aggregated specialist engagement.}
For any example $x$, we summarize specialist engagement via a signed aggregation
\begin{equation}
A_{\text{spec}}(x)=
\frac{1}{\sum_{(\ell,h)\in\mathcal{H}_{\text{spec}}}|\rho_{\ell,h}|}
\sum_{(\ell,h)\in\mathcal{H}_{\text{spec}}}\rho_{\ell,h}\, a_{\ell,h}(x),
\label{eq:Aspec}
\end{equation}
and use $A_{\text{spec}}(x)$ as an instance-level listening indicator.

\noindent\textbf{Validation protocol.}
We select $\mathcal{H}_{\text{spec}}$ using only $\mathcal{D}_{\text{cal}}$, and report all analyses on a disjoint evaluation split. We perform two sanity checks: (i) $A_{\text{spec}}(x)$ is predictive of correctness (e.g., measured by AUC) and outperforms matched random-head baselines; and (ii) $A_{\text{spec}}(x)$ is higher on examples where the model’s prediction changes between the audio-conditioned and audio-ablated runs than on examples where it does not.

\subsection{Layer-Guided Steering}
\label{sec:steering}

Let $\mathcal{H}_{\text{spec}}=\{(\ell_1,h_1),\ldots,(\ell_K,h_K)\}$ be the specialist head set and define the specialist layer set
\begin{equation}
\mathcal{L}=\{\ell:\exists h \text{ such that } (\ell,h)\in\mathcal{H}_{\text{spec}}\}.
\end{equation}
For each $\ell\in\mathcal{L}$, let $n_\ell=\big|\{h:(\ell,h)\in\mathcal{H}_{\text{spec}}\}\big|$ and set $w_\ell=n_\ell/K$ (so $\sum_{\ell\in\mathcal{L}} w_\ell = 1$).

\noindent\textbf{Steering direction (layer-localized).}
For input $x$, we run two forward passes ($x^{\text{aud}}$ and $x^{\text{sil}}$) and extract residual-stream states $\mathbf{h}^{\text{aud}}_\ell(x),\mathbf{h}^{\text{sil}}_\ell(x)$ at the final prompt position $i_{\text{final}}$. We define
\begin{equation}
\mathbf{s}(x)=\sum_{\ell\in\mathcal{L}} w_\ell\Big(\mathbf{h}^{\text{aud}}_\ell(x)-\mathbf{h}^{\text{sil}}_\ell(x)\Big).
\label{eq:sx}
\end{equation}

\noindent\textbf{Inference-time steering.}
We steer the final-layer representation by
\begin{equation}
\mathbf{h}^*(x)=\mathbf{h}^{\text{aud}}_{\text{final}}(x)+\beta\,\mathbf{s}(x),
\label{eq:steer}
\end{equation}
and compute predictions from $\mathbf{h}^*(x)$ via the language modeling head.

\subsection{Head-Level Steering}
\label{sec:head_steering}

To test whether the gains require layer-localized steering, we also consider a direct head-level intervention baseline.

\noindent\textbf{Per-head attention output.}
Let $\mathbf{u}_{\ell,h}(x)\in\mathbb{R}^{d_h}$ denote the output of attention head $h$ at layer $\ell$ (at position $i_{\text{final}}$) \emph{before} the attention output projection $W_O^{(\ell)}$.
We define the per-head audio-specific delta:
\begin{equation}
\Delta \mathbf{u}_{\ell,h}(x)=\mathbf{u}^{\text{aud}}_{\ell,h}(x)-\mathbf{u}^{\text{sil}}_{\ell,h}(x).
\end{equation}

\noindent\textbf{Mapping to the residual stream.}
Treating $\Delta \mathbf{u}_{\ell,h}(x)$ as a column vector, we map it to the residual-stream space via the corresponding slice of $W_O^{(\ell)}$:
\begin{equation}
\Delta \mathbf{c}_{\ell,h}(x)= W_O^{(\ell)}[:,\,h d_h:(h{+}1)d_h]\;\Delta \mathbf{u}_{\ell,h}(x)\in\mathbb{R}^{d_{\text{model}}}.
\end{equation}

\noindent\textbf{Intervention.}
Let $\tilde{\mathbf{h}}_{\ell}(x)$ denote the hidden state at position $i_{\text{final}}$ immediately after the attention sublayer (and before the MLP) in layer $\ell$.
Define $\mathcal{H}_{\text{spec}}(\ell)=\{h:(\ell,h)\in\mathcal{H}_{\text{spec}}\}$.
For each $\ell\in\mathcal{L}$, we add the head-level intervention to the residual stream at the output of the attention sublayer:
\begin{equation}
\tilde{\mathbf{h}}^*_{\ell}(x)=\tilde{\mathbf{h}}^{\text{aud}}_{\ell}(x)+\beta \cdot
\frac{1}{|\mathcal{H}_{\text{spec}}(\ell)|}\sum_{h\in\mathcal{H}_{\text{spec}}(\ell)}\Delta \mathbf{c}_{\ell,h}(x),
\end{equation}
and then continue the forward pass normally to obtain the final prediction.

\section{Experimental Setup}
\label{sec:exp_setup}
We evaluate on the Massive Multi-Task Audio Understanding (MMAU) benchmark \cite{sakshi2024mmau}, which consists of audio clips paired with multiple-choice questions spanning three domains: speech, environmental sound, and music, and covers 27 skills with standardized splits. We report accuracy on the labeled MMAU test-mini split (1{,}000 examples) and also break down results by domain (speech/sound/music).

\textbf{Baselines and interventions.}
We evaluate: (i) \emph{no intervention}; (ii) \emph{best single layer} audio--silence steering; (iii) \emph{head-level steering} using either specialist heads or a same-size random-head control (via per-head outputs mapped through $W_O^{(\ell)}$); (iv) \emph{matched random-head controls} for layer-guided steering (same $K$ and identical procedure); and (v) \emph{specialist-guided (head-guided layer) steering}, which aggregates audio--silence residual-state differences over layers containing specialist heads, weighted by specialist density ($w_\ell=n_\ell/K$).

\textbf{Models.}
We study two Qwen-based LALMs: Qwen2-Audio-7B-Instruct \cite{chu2024qwen2audio} and R1-AQA \cite{li2025r1aqa}, an RL-optimized audio question-answering model built on the Qwen backbone. Both models ingest audio as audio-conditioned tokens within the LLM token sequence, enabling joint audio--text inference via standard self-attention. We run all methods in a multiple-choice setting by selecting the option with the highest next-token logit for its label.

\textbf{Evaluation protocol.}
We evaluate MMAU in a 4-way multiple-choice setting. For each question with options (A--D), we score each option by the next-token logit of its label at the final prompt position $i_{\text{final}}$ (prior to generation), and predict the highest-scoring option. We report accuracy (the fraction of questions where the prediction matches the ground-truth label). Statistical significance of paired comparisons is assessed with McNemar's test. We select the steering strength $\beta$ on a held-out calibration split $\mathcal{D}_{\text{cal}}$, and report final results on the MMAU test-mini split, which is never used for hyperparameter tuning.

\textbf{Specialist head selection.}
We extract attention weights from all $L\times H$ heads on the calibration split $\mathcal{D}_{\text{cal}}$ (Qwen2-Audio-7B-Instruct and R1-AQA: $L{=}32$, $H{=}32$, i.e., $1024$ heads; both are based on a Qwen-7B backbone). For each head, we compute $a_{\ell,h}(x)$ (Eq.~\ref{eq:audio_attention}) and its correlation $\rho_{\ell,h}$ with correctness, selecting the top-$K$ heads by $|\rho_{\ell,h}|$ (we use $K{=}20$).

\textbf{Two-pass runs and activation caching.}
All steering variants use two forward passes per example: an audio-conditioned pass and a matched-duration silence pass.
For efficiency, we cache residual-stream states $\mathbf{h}_\ell(x)$ at position $i_{\text{final}}$ for all layers on $\mathcal{D}_{\text{cal}}$; for the head-level baseline, we additionally cache per-head attention outputs $\mathbf{u}_{\ell,h}(x)$ (before $W_O^{(\ell)}$) at the same position.
We later restrict computation to the specialist layer set $\mathcal{L}$ induced by the discovered heads (Section~\ref{sec:steering}).


\section{Results}

\textbf{Listening signal.}
The specialist listening score $A_{\text{spec}}(x)$ predicts correctness and substantially exceeds matched random-head controls. It also increases on examples where the predicted option changes between the audio-conditioned and audio-ablated runs ($p<0.001$), indicating that it tracks when audio affects the model's decision.

\textbf{Accuracy gains.}
Head-guided layer steering improves MMAU test-mini accuracy from $49.20\%\!\rightarrow\!57.25\%$ on Qwen2-Audio (+8.05 pp) and from $64.50\%\!\rightarrow\!69.40\%$ on R1-AQA (+4.90 pp), outperforming the best single-layer baseline (Table~\ref{tab:main_results}). A head-level specialist baseline yields non-trivial gains but remains weaker than layer-guided steering (Table~\ref{tab:main_results}).

\begin{table}[h]
\centering
\setlength{\tabcolsep}{4pt} 
\caption{Accuracy (\%) on MMAU test-mini (1{,}000 examples). Head-guided layer steering outperforms baselines.}
\label{tab:main_results}
\begin{tabular}{lcc}
\toprule
Method & Qwen2-Audio & R1-AQA \\
\midrule
Baseline (no intervention) & 49.20 & 64.50 \\
\midrule
Head-level steering (random heads) & 49.42 & 64.02 \\
Best single layer steering & 53.82 & 65.21 \\
Head-level steering (specialist) & 54.30 & 67.80 \\
\midrule
\textbf{Head-guided layer steering} & \textbf{57.25} & \textbf{69.40} \\
\bottomrule
\end{tabular}
\end{table}

\begin{table}[!htbp]
\centering
\setlength{\tabcolsep}{3.5pt} 
\caption{Accuracy (\%) by domain on MMAU test-mini; Overall reports gain in percentage points (pp).}
\label{tab:domain_breakdown}
\begin{tabular}{lccccc}
\toprule
& \multicolumn{2}{c}{Qwen2-Audio} & \multicolumn{2}{c}{R1-AQA} \\
\cmidrule(lr){2-3} \cmidrule(lr){4-5}
Domain & Baseline & Specialists & Baseline & Specialists \\
\midrule
Speech & 42.04 & 56.14  & 57.36 & 60.68  \\
Sound  & 54.95 & 59.84  & 69.37 & 76.89  \\
Music  & 50.98 & 56.07  & 66.77 & 70.67 \\
\midrule
\textbf{Overall} & \textbf{49.20} & \textbf{57.25 (+8.05)} & \textbf{64.50} & \textbf{69.40 (+4.90)} \\
\bottomrule
\end{tabular}
\end{table}

\begin{table}[h]
\centering
\caption{Layer-guided steering with specialist-selected heads vs.\ matched random-head sets: improvement over baseline (pp). Random heads induce a layer set $\mathcal{L}_{\text{rand}}$ using the same procedure as specialists.}
\label{tab:baselines}
\begin{tabular}{lccccc}
\toprule
& \multicolumn{2}{c}{Qwen2-Audio} & \multicolumn{2}{c}{R1-AQA} \\
\cmidrule(lr){2-3} \cmidrule(lr){4-5}
K & Random & Specialists & Random & Specialists \\
\midrule
10 & +3.1\% & +7.2\% & +1.0\% & +4.7\% \\
20 & +2.2\% & +8.0\% & +2.0\% & +4.9\% \\
30 & +3.1\% & +7.4\% & +1.3\% & +4.6\% \\
\bottomrule
\end{tabular}
\end{table}

\textbf{Domain breakdown.}
Improvements are consistent across domains (Table~\ref{tab:domain_breakdown}). On Qwen2-Audio, gains are largest for \textsc{Speech} (+14.1 pp), followed by \textsc{Sound} (+4.9 pp) and \textsc{Music} (+5.1 pp). On R1-AQA, gains are largest for \textsc{Sound} (+7.5 pp), with smaller improvements on \textsc{Speech} (+3.3 pp) and \textsc{Music} (+3.9 pp).

\textbf{Selection and sensitivity.}
To isolate the effect of specialist selection, we compare our layer-guided steering to a matched control where we sample $K$ heads uniformly at random, induce the corresponding layer set $\mathcal{L}_{\text{rand}}$ (and weights $w_\ell$) using the same procedure as for specialists, and apply the identical layer-guided steering intervention. Matched random-head sets yield much smaller improvements than specialists across $K$ (Table~\ref{tab:baselines}), showing that the gains are driven by the discovered heads rather than by steering arbitrary layers. Figure~\ref{fig:steering_grid} shows a consistent operating regime: performance peaks at moderate $\beta$ (with $K\!\approx\!20$ typically near-optimal) and degrades for overly large $\beta$, suggesting over-steering. The induced specialist layer set remains sparse as $K$ increases (Table~\ref{tab:layers}).

\begin{table}[!htbp]
\centering
\caption{Induced specialist layer set $\mathcal{L}$ as a function of $K$ (see Section~\ref{sec:steering}). ``Added layers'' lists layers newly included in $\mathcal{L}$ when increasing $K$; $|\mathcal{L}|$ is the resulting set size.}
\label{tab:layers}
\begin{tabular}{lccccc}
\toprule
& \multicolumn{2}{c}{Qwen2-Audio} & \multicolumn{2}{c}{R1-AQA} \\
\cmidrule(lr){2-3} \cmidrule(lr){4-5}
$K$ & Added layers & $|\mathcal{L}|$ & Added layers & $|\mathcal{L}|$ \\
\midrule
10 & 19, 21, 22, 23 & 4  & 16, 17, 18, 19, 22 & 5  \\
20 & 14, 18, 20, 24, 26 & 9  & 20, 24, 27, 28, 30 & 10 \\
30 & 25, 27 & 11 & 26 & 11 \\
\bottomrule
\end{tabular}
\end{table}

\section{Discussion}
Our results highlight mechanistic interpretability as a practical tool for understanding and improving audio-language models. Head-level attention analysis yields an instance-level indicator of audio engagement and localizes a sparse set of specialist heads where audio-relevant computation concentrates. Using an audio--silence counterfactual, we show that selectively intervening in these layers can amplify the model’s \emph{audio effect} and produce consistent accuracy gains (up to +8 pp on MMAU) without parameter updates. Overall, this suggests that text dominance in LALMs is a diagnosable and steerable failure mode, and that interpretability can provide actionable localization signals for building more reliably grounded multimodal systems.

\newpage

\bibliographystyle{IEEEtran}
\bibliography{mybib}

\end{document}